**Title:** **Second-generation *p*-values: improved rigor, reproducibility, & transparency in statistical analyses**

**Authors:** Jeffrey D. Blume[1*] PhD, Lucy D'Agostino McGowan[2] MS, William D. Dupont[3] PhD, Robert A. Greevy[1] Jr. PhD

**Affiliations:**

[1]Associate Professor, Department of Biostatistics, Vanderbilt University School of Medicine, Nashville TN
[2]PhD Candidate, Department of Biostatistics, Vanderbilt University School of Medicine, Nashville TN
[3]Professor, Department of Biostatistics, Vanderbilt University School of Medicine, Nashville TN

[*]Correspondence to: j.blume@vanderbilt.edu

**Abstract:** Verifying that a statistically significant result is scientifically meaningful is not only good scientific practice, it is a natural way to control the Type I error rate. Here we introduce a novel extension of the *p*-value – a second-generation *p*-value $(p_\delta)$ – that formally accounts for scientific relevance and leverages this natural Type I Error control. The approach relies on a pre-specified interval null hypothesis that represents the collection of effect sizes that are scientifically uninteresting or are practically null. The second-generation *p*-value is the proportion of data-supported hypotheses that are also null hypotheses. As such, second-generation *p*-values indicate when the data are compatible with null hypotheses $(p_\delta = 1)$, or with alternative hypotheses $(p_\delta = 0)$, or when the data are inconclusive $(0 < p_\delta < 1)$. Moreover, second-generation *p*-values provide a proper scientific adjustment for multiple comparisons and reduce false discovery rates. This is an advance for environments rich in data, where traditional *p*-value adjustments are needlessly punitive. Second-generation *p*-values promote transparency, rigor and reproducibility of scientific results by *a priori* specifying which candidate hypotheses are practically meaningful and by providing a more reliable statistical summary of when the data are compatible with alternative or null hypotheses.

**One Sentence Summary:** Second generation *p*-values preserve the simplicity that has made *p*-values popular while resolving critical flaws that lead to misinterpretation of data, distraction by trivial effects, or unreproducible assessments of data.

**Main Text:**

### 1. Introduction

*P*-values abound in the scientific literature. They have become the researcher's essential tool for summarizing when the data are incompatible with the null hypothesis. Although *p*-values are widely recognized as imperfect tools for this task, the impact of their flaws on scientific inference remains hotly debated (1-5). The debate over the proper use and interpretation of *p*-values has stymied and divided the statistical community (6-14). Recurring themes include the difference between statistical and scientific significance, the routine misinterpretation of non-significant *p*-values, the unrealistic nature of a point null hypothesis, and the challenges with multiple comparisons. With no widely-accepted alternative to promote, statisticians are left to tweak the manner in which *p*-values are applied and interpreted (11,12). Some have even suggested that the problem lies with instruction: *p*-values are fine, they are just widely misused (15,16). After a century of widespread adoption in science, with their flaws and advantages well-known, it is time for an upgrade.

The purpose of this paper is to introduce a novel and intuitive extension that better serves the *p*-value's intended purpose. We call this upgrade a second-generation *p*-value. Second-generation *p*-values are easy to compute and interpret. They offer improved inferential capability, e.g. it is now possible for the data to indicate support for the null hypothesis. They control the Type I error naturally, forcing it to zero as the sample size grows. This, in turn, offsets Type I Error inflation that results from multiple comparisons or multiple examinations of accumulating data. Findings identified by second-generation *p*-values are less likely to be false discoveries than findings identified by classical *p*-values. Consequently, second-generation *p*-values do not require ad-hoc adjustments to provide strict error control and this improves power in studies with massive multiple comparisons. They also implicitly codify good research practice: the smallest effect size of scientific relevance must now be specified before looking at results. This prevents the inevitable rationalization that accompanies the post-hoc interpretation of mediocre results that have been deemed statistically significant. This singular change alone will improve rigor and reproducibility across science.

Our examples (Section 3) were selected from a wide range of contexts to highlight the broad utility of this new tool. We will not dwell on the well-known drawbacks of classical *p*-values (11,12,13,14). The frequency properties of second-generation *p*-values are the same or better than traditional *p*-values. These technical details, along with supplementary exposition, can be found in the supplementary materials. A distinguishing feature of second-generation *p*-values is that they are intended as summary statistics that indicate when a study has met its a priori defined endpoint: the observed data support only alternative hypotheses or only null hypotheses.

Given the complexity surrounding the interpretation and computation of *p*-values, and the plethora of ad-hoc statistical adjustments for them, the reader is forgiven for any pre-emptive statistical fatigue, pessimism, or skepticism. After all, every statistical adjustment for multiple comparisons boils down to nothing more than ranking the *p*-values and picking a cutoff to determine significance. While each method offers its own preferred cut-off, the core value judgement – the ranking – remains the same. Second generation *p*-values, however, change that ranking; they favor results that are both scientifically relevant and statistically significant. For example, Section 3 presents an application where a Bonferroni correction yields 264 genes of interest from a study of 7128 candidate genes where 2028 had an unadjusted *p*-value of 0.05 or less. An application of the second-generation *p*-value also yields 264 gene findings (their second-generation *p*-value is 0), ensuring the same Type I Error control. However, 82 (31%) of those genes fail to meet the Bonferroni criteria. The difference is both fascinating and striking, and is due to the second-generation *p*-value's preference for scientific relevance (which in this case amounts to a preference for clinically relevant fold changes in expression levels).

## 1.1 Illustration of approach

The top diagram of Figure 1 depicts an estimated effect, typically the best supported hypothesis $\hat{H}$, its 95% confidence interval (CI), and the traditional point null hypothesis, $H_0$. The CI contains all the effect sizes that are supported by the data at the 95% level; we will refer to it as the set of data-supported hypotheses. If the null hypothesis is well outside of the interval, the *p*-value is very small or near zero. If the CI just barely excludes the null hypothesis, the *p*-value will be slightly less than 0.05. When the CI contains the null hypothesis, the *p*-value will be larger than 0.05. The *p*-value grows to 1 as the null hypothesis approaches the center of the CI.

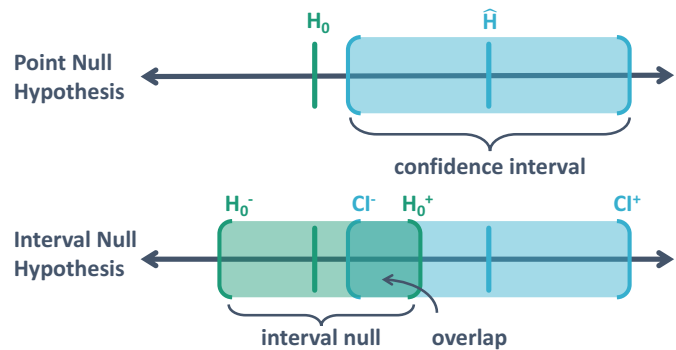

**Figure 1:** *Illustration of a point null hypothesis, $H_0$; the estimated effect that is the best supported hypothesis, $\widehat{H}$; the 95% confidence interval (CI) for the estimated effect $[CI^-, CI^+]$; and the interval null hypothesis $[H_0^-, H_0^+]$.*

Now imagine that the null hypothesis is a contiguous set – an interval – rather than just a single point, as depicted in the bottom diagram of Figure 1. The interval null is the set of effects that are indistinguishable from the null hypothesis, due to limited precision or practicality. For example, the null hypothesis "no age difference" might be re-framed as "no age difference of more than 365 days", the latter being what we really mean when we say two people are the same age (e.g., they are both 45). An interval null always exists, even if it is narrow.

When a 95% CI is entirely contained within the null interval, the data support only null hypotheses (this is the traditional benchmark for showing statistical equivalence). When the CI and null set do not overlap, the data are said to be incompatible with the null. Lastly, when the null set and confidence interval partially intersect, the data are inconclusive. Thus, the degree of overlap conveys how compatible the data are with the null premise. The second-generation *p*-value is the fraction of overlap multiplied by a small-sample correction factor. We define it formally in Section 2. *In a very real sense, the second-generation p-value is nothing more than the codification of today's standards for good scientific and statistical practice.*

### 1.2 Interval vs. point null hypothesis

The formal acknowledgement of an interval null hypothesis has important consequences for how statistical methods are applied and understood in science. Any point hypothesis, while mathematically convenient, represents a statistical hypothesis so precise it can never be confirmed with finite data. For null hypotheses, this high level of specificity can complicate inference. For example, when a point null hypothesis is "rejected", it can be the case that there are other point hypotheses, practically indistinguishable from the point null, that remain well supported by the data. The solution is to use an interval null hypothesis. These are constructed by incorporating information about the scientific context – such as inherent limits on measurement precision, clinical significance, or scientific significance – into statistical hypotheses that are stated *a priori*. The tag 'scientifically relevant' or 'clinically significant' is reserved for effects (hypotheses) that are non-trivial and meaningful, i.e., beyond the interval null hypothesis. Measurement precision can always be used to establish a lower bound on scientifically relevant hypotheses, as it makes little sense to ponder effect sizes that are smaller than the detectable limit. The interval null should contain, in addition to the precise point null hypothesis, all other point hypotheses that are practically null and would maintain the scientific null premise. While the point null may be numerically distinct, all the hypotheses in the interval null are considered scientifically equivalent to the null premise.

## 2. Definition and computation

### 2.1 Formula

Let $I$ represent the interval of hypotheses for a scalar parameter that are best supported by the data – an unadjusted 95% CI for example – and let $H_0$ represent the interval null hypothesis. If $I = [a, b]$ where $a < b$ are real numbers, then its length is $|I| = b - a$. The second-generation *p*-value, denoted by $p_\delta$, is defined as

$$p_\delta = \frac{|I \cap H_0|}{|I|} \times \max\left\{\frac{|I|}{2|H_0|}, 1\right\} \tag{1}$$

where $I \cap H_0$ is the intersection, or overlap, between intervals $I$ and $H_0$. The subscript $\delta$ signals the reliance of the second-generation *p*-value on the interval null. Numerically, $\delta$ represents the half-width of the interval null hypothesis. The value of $\delta$ is driven by scientific context and should be specified prior to conducting the experiment. The first term in equation (1) is the fraction of best supported hypotheses that are also null hypotheses. The second term is a small-sample correction factor, which forces the second-generation *p*-value to indicate inconclusiveness when the observed precision is insufficient to permit valid scientific inferences about the null hypotheses.

As described here, $p_\delta$ is the length of the intersection between the two intervals, divided by the length of the interval estimate, multiplied by the correction factor. When the interval estimate is sufficiently precise, defined here as when $|I| < 2|H_0|$, the second-generation *p*-value is just the overlap fraction, $|I \cap H_0|/|I|$. When the interval estimate is very wide, $|I| > 2|H_0|$, the second-generation *p*-value reduces to $0.5 \times |I \cap H_0|/|H_0|$, which is bounded by ½.

Definition (1) readily extends to multiple dimensions to accommodate parameter vectors. In that case, $|I|$ would represent an area or volume. Neither interval is required to be symmetric or of finite length. Although the vast majority of intervals in the literature are 'two-sided' with finite length, pathologies are possible when neither interval has finite length. If intervals $I$ and $H_0$ overlap but neither has finite length, e.g., overlapping one-sided intervals, the second-generation *p*-value will be zero or one, depending respectively on whether $|I \cap H_0|$ is finite or infinite (SR1). Note that any interval estimate, from a likelihood support interval to a Bayesian credible interval, could be used in place of the 95% confidence interval (more on this point in Section 2.6). Inferential denomination, computational ease, and desired frequency properties could inform this choice.

### 2.2 A simple example

When measuring systolic blood pressure (SBP), the currently accepted recommendation is to report SBP to the nearest 2 mmHg when using analog or mercury devices (17,18). Blood pressure dials indicate even numbers and changes less than 2 mmHg are not clinically actionable. Figure 2 and Table 1 display results from 8 mock studies.

**Figure 2**: *Forest plot of mock results from 8 studies of systolic blood pressure. Here the point null is 146 mmHg, indicated by the vertical dashed line, with an indifference zone, or interval null hypothesis, from 144 mmHg to 148 mmHg shaded in blue-grey.*

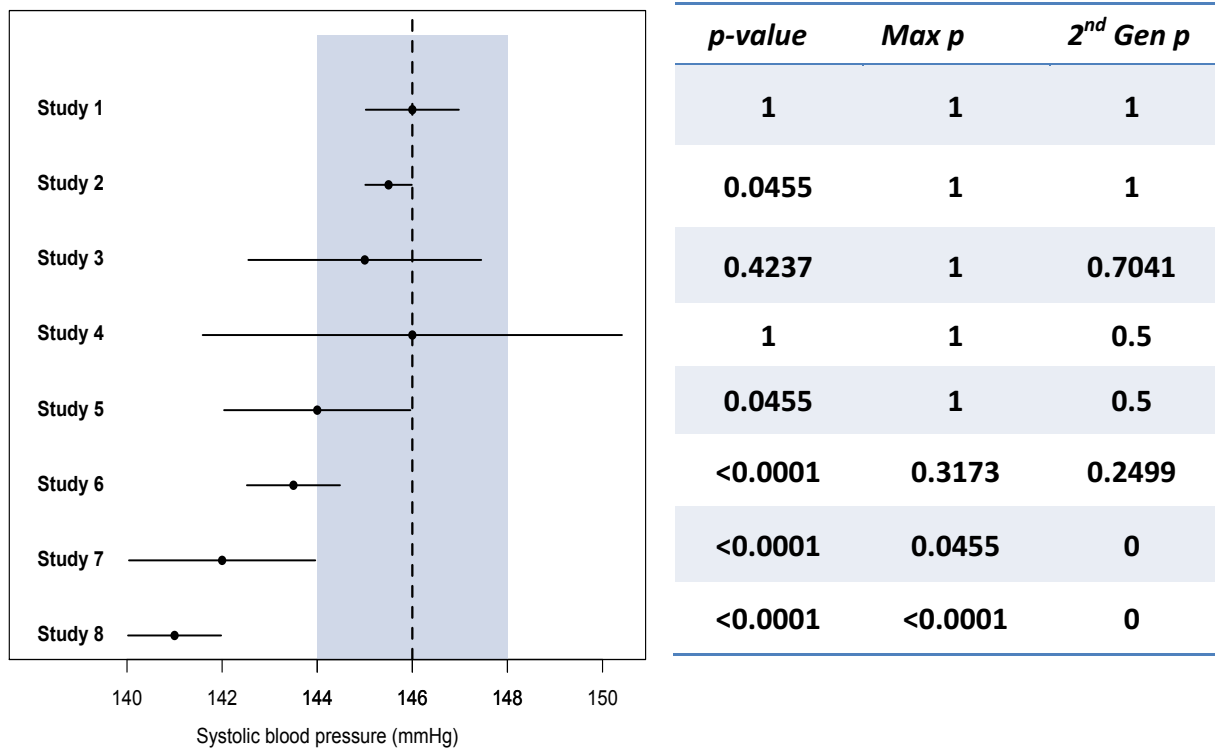

**Table 1:** *Mock results from 8 studies of systolic blood pressure.*

| Study | Mean (SE) | 95% CI Lower | 95% CI Upper | 2$^{nd}$ Gen. $p$-value ($p_\delta$) | Maximum $p$-value | Traditional $p$-value |
|---|---|---|---|---|---|---|
| 1 | 146 (0.5) | 145.02 | 146.98 | 1 | 1 | 1 |
| 2 | 145.5 (0.25) | 145.01 | 145.99 | 1 | 1 | 0.0455 |
| 3 | 145 (1.25) | 142.55 | 147.45 | 0.7041 | 1 | 0.4237 |
| 4 | 146 (2.25) | 141.59 | 150.41 | 0.5 | 1 | 1 |
| 5 | 144 (1) | 142.04 | 145.96 | 0.5 | 1 | 0.0455 |
| 6 | 143.5 (0.5) | 142.52 | 144.48 | 0.2449 | 0.3173 | <0.0001 |
| 7 | 142 (1) | 140.04 | 143.96 | 0 | 0.0455 | <0.0001 |
| 8 | 141 (0.5) | 140.02 | 141.98 | 0 | <0.0001 | <0.0001 |

Table 1 reports the mean SBP, the 95% CI, the second-generation $p$-value based on $\delta = 2$ mmHg, the maximum $p$-value over all possible null hypotheses in the range $H_0: 144 \leq \mu \leq 148$, and the traditional $p$-value for testing $H_0: \mu = 146$.

For example, in Study 3, the upper confidence bound is 142.55 and the lower confidence bound is 147.45. The interval null, $\mu_0 \pm \delta = [144, 148]$, has length $|H_0| = 2\delta = 4$ mmHg. So $p_\delta$ is

$$p_\delta = \frac{(147.45 - 144)}{(147.45 - 142.55)}(1) = 0.7041 \tag{2}$$

A $p_\delta$ of 0.7 means the data are inconclusive with a slight or weak favoring of null hypotheses. Intuitively, it means that 70% of the data-supported hypotheses are null hypotheses. Study 4 is an example where the correction factor comes into play. The length of the confidence interval, 8.82, is more than twice the length of the null interval, 4. Therefore, $p_\delta$ is:

$$p_\delta = \frac{(148 - 144)}{(150.41 - 141.59)} \frac{(150.41 - 141.59)}{2(148 - 144)} = 0.5 \tag{3}$$

A $p_\delta$ of 0.5 means the data are strictly inconclusive. The role of the correction factor is discussed in Section 2.5.

The second-generation $p$-values best describe Figure 2. Study 1 is an example where the data only support null effects. Study 2 demonstrates the paradox of statistical and scientific significance. Both $p_\delta$, and the maximum $p$-value seem to account for this paradox. However, $p_\delta$ has more desirable attributes in other circumstances, such as in Studies 3 to 6. Study 4 has a traditional $p$-value and maximum $p$-value of 1, but the data are clearly inconclusive. The second-generation $p$-value reflects this, with a value of 0.5, allowing for a more nuanced interpretation of 'inconclusive'. In studies where the point estimate falls in the null range, but the confidence interval extends beyond (e.g., studies 3, 4, and 5) the maximum *p*-value does not properly convey what the data are saying, and the traditional *p*-value does not account for the range of null values, or the precision of the measurement tool. Studies 7 and 8 illustrate the case when the confidence interval is fully beyond the null space.

### 2.3 Interpretation

Consider a general interval null hypothesis, say $H_0: \mu_{0a} \leq \mu \leq \mu_{0b}$. The second-generation *p*-value, $p_\delta$, has the following properties:

1. $p_\delta$ is a proportion; a number between zero and one, inclusive.

2. $p_\delta$ is the fraction of data-supported hypotheses that are null hypotheses and therefore compatible with the null premise.

3. When $p_\delta = 0$, the data only support hypotheses that are scientifically or clinically meaningful, i.e., those that are meaningful alternative hypotheses.

4. When $p_\delta = 1$, the data only support null hypotheses, i.e., those that are *not* scientifically or clinically meaningful.

5. When $p_\delta \approx 1/2$, the data are strictly inconclusive. The degree of inconclusiveness is represented by $p_\delta$ itself. For example, $p_\delta = 1/8$ and $p_\delta = 7/8$ both represent the same degree of inconclusiveness, but the balance between alternative and null hypotheses is reversed.

6. $p_\delta$ has improved error rate control when the interval estimate $I$ is a $100(1-\alpha)\%$ CI. Under any null hypothesis within the interval null, the probability of observing $p_\delta = 0$ is less than or equal to $\alpha$. This probability converges to zero as the sample size grows. Under any hypotheses beyond the interval null, the probability of observing $p_\delta = 0$ converges to one as the sample size grows.

The interpretation of $p_\delta$ may appear similar to that of a posterior probability of the null hypothesis. However, $p_\delta$ is strictly not a posterior probability and it is not an estimate of the probability of the null hypothesis. Rather, $p_\delta$ is simply the observed fraction of data-supported hypotheses that are null hypotheses; it is a descriptive statistic – a simple proportion. Essential elements of a proper posterior computation, namely knowledge that some hypotheses are better supported than others and the degree to which some hypotheses inside the interval null should be favored over others, are not needed. Specification of the latter is controversial because it does not depend on the data at hand. Second-generation *p*-values are descriptive; they indicate when the study has generated data that rule out null or alternative hypotheses.

The third property does not hold for traditional *p*-values; this is why *statistical significance is not scientific significance*. The fourth property is strictly false for classical *p*-values; a *p*-value larger than the Type I Error probability $\alpha$ is considered inconclusive. Large *p*-values never "support" the null hypothesis; they just indicate the lack of strong evidence against it. This is the *absence of evidence is not evidence of absence* conundrum. The fifth property is also not strictly true for traditional *p*-values; when non-significant, the *p*-value is to be interpreted as inconclusive despite the temptation to do otherwise. A welcome feature of second-generation *p*-values is that they distinguish between data that are inconclusive ($0 < p_\delta < 1$) and data that are compatible with null hypotheses ($p_\delta = 1$). This ability is sorely needed in practice.

The sixth property is a major improvement. Unlike its predecessor, the second-generation *p*-value converges to zero or one as the sample size grows, which means the procedure is inferentially consistent in the limit (SR2). Why this happens is interesting and intuitive (SR3). The take-home message is that their frequency properties are no worse than those of the interval estimates upon which the second-generation *p*-value is based, and are often improved in moderate to large samples. A detailed examination of the frequency properties of second-generation *p*-values is included in the supplement (SR10-SR18).

### 2.4 The Delta Gap

It can be helpful to have a way of ranking two studies that both have second-generation *p*-values of zero ($p_\delta = 0$). One way to do this is to use the *delta-gap*, which is the distance between the intervals in $\delta$ units. Recall that $\delta$ is the half-width of the interval null hypothesis. If the CI was shifted to the right of the null interval, the delta-gap would be $(\text{CI}^- - \text{H}_0^+)/\delta$. Scaling by $\delta$ makes it unit free. The delta-gap ranking favors extremes in effect size, which is a logical complement to the second-generation *p*-value. Remember that second-generation *p*-values provide a quick and easy marker of when a study reaches a natural endpoint, i.e., when the data are compatible with only null hypotheses ($p_\delta = 1$) or only alternative hypotheses ($p_\delta = 0$). Two studies with equal second-generation *p*-values do not necessarily represent equal amounts of statistical evidence. For example, their likelihood functions may not be proportional. The same is true of classical *p*-values, of course (11,12).

### 2.5 Role of small-sample correction factor

The small-sample correction factor comes into play when the intervals overlap and the range of data supported hypotheses is more than twice that of the indifference zone, i.e., when $|I| > 2|H_0|$. Because the width of the interval estimate $I$ will shrink as the sample size grows, the correction factor comes into play more often in small samples.

The second-generation *p*-value is based on the proportion of data-supported hypotheses that are null, or practically null, hypotheses. This is written as $|I \cap H_0|/|I|$. When the data are very imprecise, this proportion alone can be misleading. To see this, consider the case when $|I| \gg |H_0|$ and the two intervals completely overlap. The proportion alone would be small, indicating that the data favor alternative hypotheses even though every possible null hypothesis is just as well supported. Clearly the data are inconclusive, and the correction factor resets the proportion to ½ to indicate this.

The correction factor allows second-generation *p*-values greater than ½ to reliably represent degrees of compatibility with the null hypotheses. When the correction factor is applicable, the second-generation *p*-value is bounded above by ½, which is achievable only when the null interval is entirely contained within the interval estimate. When the intervals fail to intersect, the relative degree of precision is immaterial and $p_\delta = 0$ because there are no hypotheses in common.

How often will the correction factor come into play? It depends on a number of factors. For planned experiments, a general benchmark is that the correction factor plays a role when the power to detect the smallest meaningful hypothesis drops below 16% (SR4). That is, it comes into play only for studies that are severely underpowered to detect meaningful effect sizes.

### 2.6 Choosing an interval estimate

In this paper, we choose 95% CIs for the interval estimate $I$. However, the definition of a second-generation *p*-value is not so exclusive. Any interval estimate, from a likelihood support interval to a Bayesian credible interval, could be used. Different interval estimates will impart different frequency properties to the second-generation $p$-value. The interpretation and usage of $p_\delta$ would remain unchanged as long as $I$ represented, in some sense, a set of best supported hypotheses. Our preference is to use a 1/8 likelihood support interval (SI). SIs have a well-established evidential interpretation, are not dependent on the sample space, and are otherwise well aligned with the $p$-value's *raison d'être* (21,22). A 1/8 likelihood support interval – the benchmark for moderate strength of evidence – is equivalent to an unadjusted 95.9% CI in many regular cases, so their distinction is largely interpretative (11,21). Because confidence intervals have near universal familiarity, and because support intervals are essentially unadjusted confidence intervals, we decided not to feature support intervals in this exposition. This choice should not be mistaken for an implicit endorsement of confidence intervals (SR5).

## 3. Applications

The first example contrasts the behavior of second-generation $p$-values with classical methods when examining differential gene expression in 7128 candidate genes. The second is a comparison of Kaplan-Meier survival curves, often done in biomarker development, where the multiple comparisons issue is hidden by design. Additional examples, for 2x2 contingency tables (SR8) and regression models (SR9), can be found in the supplement.

### 3.1 Multiple comparisons: Leukemia Microarray Study

The ALL-AML Leukemia Microarray Study (23) involved 72 leukemia participants. Forty-seven had acute lymphoblastic leukemia (ALL) and 25 had acute myeloid leukemia (AML). The study sought to identify genes with differential expression between the two types of leukemia. A microarray analysis was conducted on 7128 candidate genes and two-group t-tests were performed on the cleaned and standardized $\log_{10}$ expression levels of each gene. Here the effect scale is the difference in the logarithm of gene expression levels or fold-change (log-ratio). Typically, a fold change must be greater than 2 to draw interest, implying our interval null should be from ½ to 2, or -0.3 to 0.3 the $\log_{10}$ scale. This demarcation does not imply that fold-change is a surrogate for scientific importance. It simply means that an estimated fold-change must meet some minimum criterion to be considered interesting. It is the small effects that, when statistically significant, tend to be false discoveries (See Section 4.2).

**Figure 3:** *Display of 95% confidence intervals for gene specific fold-changes (AML vs. ALL) in the gene expression levels of patients from the Leukemia Microarray Study (23). Genes are sorted on the x-axis by classical p-value rank. Interval null hypothesis (blue-grey zone) shows all absolute fold changes between ½ and 2. Red genes have a second-generation p-value of 0, blue genes do not. Vertical dashed lines show various traditional p-value cutoffs at the 0.05 level. Genes 3252 (light blue) and 2288 (green) have a second-generation p-value of 0, while gene 350 (dark blue) has a second-generation p-value of 1.*

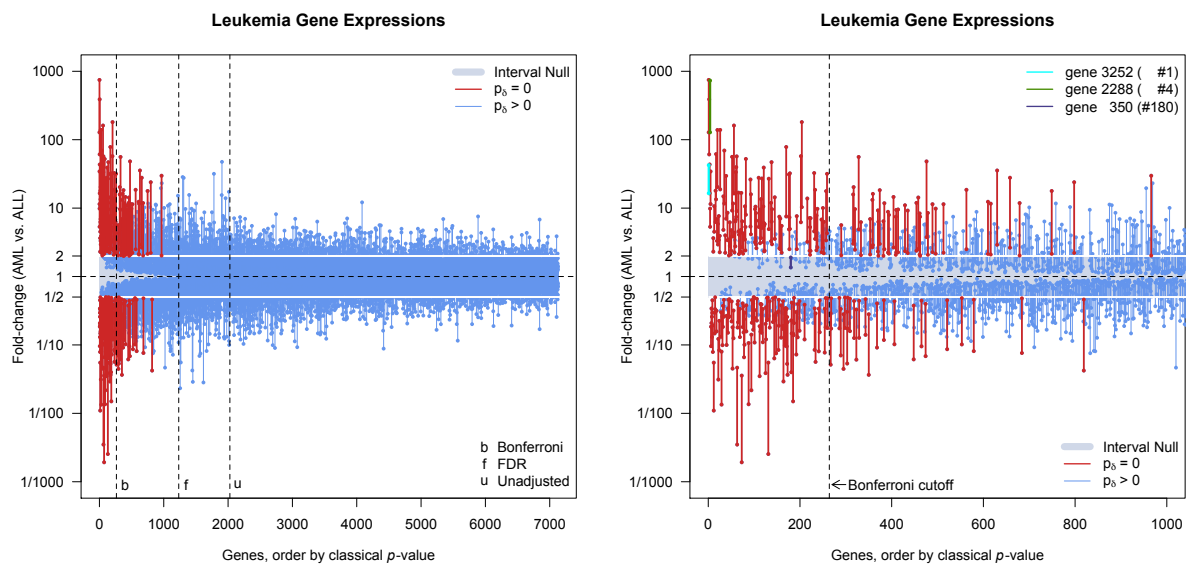

The left panel of Figure 3 displays 7128 unadjusted 95% CIs for fold change when the genes are ordered by their classical *p*-value rank. The vertical dotted lines show the Bonferroni cutoff, the empirical false discovery rate (FDR or q-value) cutoff (24), and the unadjusted *p*-value cutoff that results from $\alpha = 0.05$. Only 264 genes remain 'statistically significant' after a Bonferroni correction; 1233 have an estimated FDR less than $\alpha$; and 2028 genes have an unadjusted classical *p*-value less than $\alpha$. The indifference / null zone is shaded in a blue-grey color. The 229 CIs in red have a second-generation *p*-value of zero ($p_\delta = 0$); none of the fold changes in these CIs are between ½ and 2. Note that these findings cannot be ordered on the basis of the second-generation *p*-value alone. CIs that extend into the blue-grey zone have a $p_\delta > 0$; each of them indicates that the data support fold changes between ½ and 2. Under a global null hypothesis, the Bonferroni and the second-generation *p*-value approaches have nearly the same apparent Type I Error rate for findings: 0.037 (=264/7128) vs. 0.032 (=229/7128).

The right panel of Figure 3 displays the top 1,000 genes according to their *p*-value rank. There are 229 genes with $p_\delta = 0$ (i.e., red confidence intervals). Of these, 65 are missed by Bonferroni. The indifference zone need only be narrowed slightly, 1/1.915 to 1.915, to also have exactly 264 second-generation *p*-values that are 0. Yet even then, Bonferroni still misses 82 of them. Table 2 provides a cross-tabulation for this comparison. Consider gene #6345, whose *p*-value of 0.0033 is ranked 966[th] but has a 95% CI for fold change of 2.02 to 29.74, and gene #350, whose *p*-value of 9.02×10[-7] is ranked 180[th] but has 95% CI for fold change of 1.36 to 1.94. Bonferroni finds the second gene, where the data support only trivial fold changes, and misses the first, where the data support only meaningful fold changes. The second-generation *p*-value does just the opposite.

**Table 2:** *Cross-tabulation of second generation p-values with Bonferroni corrected p-values.*

|  | 1/2 < Fold Change < 2 ($\delta = 0.3$) | | 1/1.915 < Fold Change < 1.915 ($\delta = 0.282$) | |
| --- | --- | --- | --- | --- |
|  | $p_\delta = 0$ | $p_\delta > 0$ | $p_\delta = 0$ | $p_\delta > 0$ |
| $p_{bon} < 0.05$ | 164 | 100 | 182 | 82 |
| $p_{bon} > 0.05$ | 65 | 6799 | 82 | 6782 |
| **Total** | 229 | 6899 | 264 | 6864 |

The empirical FDR criterion could be lowered to 2.45% (slightly more than half) and still capture the same 229 genes with $p_\delta = 0$. However, it would also capture an additional 737 genes for which $p_\delta > 0$. All either overlap the null interval substantially or are contained in the null interval. As we will see in section 4.2, the actual FDR depends on a number of factors, and it does not necessarily make sense to apply the same FDR criterion to all genes. Moreover, the false discovery rate for the second-generation *p*-value is less than that for the classical *p*-value (or comparable to, depending on the significance level). All this means is that adjusting the *p*-value cutoff is not likely to help much, because it leaves the original rank ordering unchanged.

Second-generation *p*-values are selecting a fundamentally different set of candidate genes. An important point is that this set cannot be identified by standard methods where selection is based only on *p*-value ranking. To illustrate this, the supplement details what happens in the leukemia example as the indifference zone shrinks to zero (SR6). Consider also, the delta-gap ranking among the 229 genes where $p_\delta = 0$. Gene #2288, for example, has the 4[th] lowest traditional *p*-value, but has the largest a delta gap at 6 (computed on $\log_{10}$ scale). Gene #3252 has the lowest traditional *p*-value, but has a delta gap of 3 (10[th] largest). SR18 details these computations.

### 3.2 Survival

Survival data on patients with advanced lung cancer are available from the North Central Cancer Treatment Group (25). Figure 4 displays Kaplan-Meier survival curves for women (pink) and men (blue). A log-rank test (p=0.0013) indicates the curves are statistically different *somewhere*. Second-generation *p*-values tells us where. Suppose survival differences greater than 5% are of interest, implying an interval null from -0.05 to 0.05 percentage points.

**Figure 4:** *Survival in patients with advanced lung cancer from the North Central Cancer Treatment Group study. Kaplan-Meier survival curves by gender (blue for men, pink for women). Rug plot on x-axis displays second-generation p-values for the difference in survival time. Green ticks indicate incompatibility with null hypotheses; red indicate compatibility; gray indicate inconclusive results.*

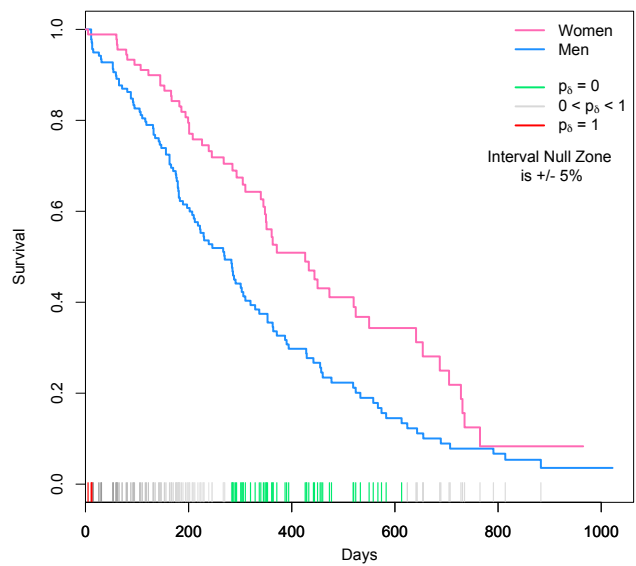

The rug plot on Figure 4 shows color coded second-generation *p*-values at every observed difference in the Kaplan-Meier curves. The ticks are green when $p_\delta = 0$ (real difference), red when $p_\delta = 1$ (no meaningful difference, if any), and shades of grey when inconclusive. It is easy to see that the curves differ by at least 5 percentage points between 300 and 600 days. Otherwise the apparent differences are inconclusive except for at the beginning when the curves are essentially the same. Remark 7 in the supplement details these calculations (SR7). There are 139 points where the curves can be compared, but this is immaterial to the second-generation *p*-value. The log-rank test avoids the implicit multiple comparisons issue, but at the cost of being non-specific. It cannot identify where the curves are different, nor is the test itself unique (26).

A common alternative approach is to use a Cox model with gender as a covariate and rely on the proportional hazards assumption. This yields an estimated hazard ratio of 1.7 (men to women) with a 95% CI of 1.23 to 2.36. Assuming that that hazard ratios between 0.9 and 1.1 are not clinically interesting, we have a $p_\delta = 0$. Hence the data, incompatible with null hypotheses, suggest a meaningful difference in lung cancer risk between men and women.

### 3.3 Additional examples and applications

The supplement contains an application of second-generation *p*-values for assessing an odds ratio in a 2x2 contingency table (SR8) and for assessing whether the data are compatible with the removal of a set of predictors from a linear regression model (SR9).

### 4. Frequency properties of second generation p-values

Second-generation *p*-values generally maintain the kind of error rate control that science has become accustomed to. Moreover, they are a more reliable inferential tool than classical *p*-values. A technical treatment of this topic is provided in supplemental remarks 10 through 18 (SR10-SR17). Here we briefly recount the key findings.

### 4.1 Behavior of second-generation *p*-values under presumed conditions

An examination of the stochastic behavior of second-generation *p*-values is revealing. Upon collecting data, a second-generation *p*-value may indicate compatibility with the alternative hypothesis ($p_\delta = 0$), compatibility with null hypotheses ($p_\delta = 1$), or inconclusiveness ($0 < p_\delta < 1$). How often these events occur, under various null and alternative hypotheses, is of interest when designing a study.

When a null hypothesis is true, we do not observe $p_\delta = 0$ often. In fact, $P(p_\delta = 0|H_0) \leq \alpha$ and often $P(p_\delta = 0|H_0) \ll \alpha$. Here $(1 - \alpha)$ denotes the coverage probability of the interval estimate. Unlike the Type I Error rate of hypothesis testing, $P(p_\delta = 0|H_0)$ shrinks to zero as the sample size grows so long as $H_0$ is in the interior of the null interval. This is partly why second-generation *p*-values are advantageous in multiple comparison settings. At the edges, the probability remains constant at $\alpha$.

When a true alternative hypothesis is outside the null interval, we do not often observe $p_\delta = 1$. This event cannot occur until the interval estimate is narrow enough to be contained by the interval null, i.e. when the sample size is large. Here too, $P(p_\delta = 1|H_1) \leq \alpha$ for any $H_1$ outside the null interval with $P(p_\delta = 1|H_1) \ll \alpha$ for $H_1$ not at the edge of the null interval. The $P(p_\delta = 1|H_1)$ also shrinks to zero as the sample size grows.

The least desired outcome is an inconclusive second-generation *p*-values, $0 < p_\delta < 1$. A common occurrence in small studies, the probability of this happening is written as $P(0 < p_\delta < 1|H)$. This probability reaches its maximum, $1 - \alpha$, when the true hypothesis $H$ is on the edge of the null interval. As the true hypothesis $H$ moves away from the edge, the probably decreases to zero. For example, the supplement shows that $P(0 < p_\delta < 1|H) \approx 0.15$ when $H$ is *three standard errors* from the edge. Hence, sample size is a means of controlling this probability. In this sense, the probability of observing an inconclusive $p_\delta$ is analogous to the Type II Error rate of hypothesis testing.

Second-generation *p*-values do sacrifice some power to characterize findings into three categories. The criteria for rejecting the null is more stringent, and inconclusive results must be separated from those supporting the null. However, the reduction in power is typically less than that caused by popular multiple comparison adjustments such as Bonferroni. As such, in contexts with massive multiple comparisons and varying standard errors, the second-generation *p*-value outperforms Bonferroni because its error rate profile is generally superior (SR21).

Although this improved inferential clarity comes with a real cost, it also yields a critical advantage: the second-generation *p*-value has a lower false discovery rate than a comparable hypothesis test. That is, second-generation *p*-values are more reliable tools than classical *p*-values.

### 4.2 Reliability of an observed second-generation *p*-value

Once data are collected and the second-generation *p*-value is computed, the long-run behavior of second-generation *p*-values becomes irrelevant. The relevant quantity is the probability that the observed results, say $p_\delta = 0$ or $1$, are mistaken. This tendency, which we will refer to as the reliability of second-generation *p*-values, is captured by the false discovery rate (FDR) $P(H_0|p_\delta = 0)$ and the false confirmation rate (FCR) $P(H_1|p_\delta = 1)$.

A straightforward application of Bayes rule allows us to compute these rates as

$$P(H_0|p_\delta = 0) = \left[1 + \frac{P(p_\delta=0|H_1)}{P(p_\delta=0|H_0)}r\right]^{-1} \text{and} \quad P(H_1|p_\delta = 1) = \left[1 + \frac{P(p_\delta=1|H_0)}{P(p_\delta=1|H_1)}\frac{1}{r}\right]^{-1} \quad (4)$$

where $r = P(H_1)/P(H_0)$. The dependence of these rates on the design probabilities $P(p_\delta = 0|H_1)$, $P(p_\delta = 0|H_0)$, $P(p_\delta = 1|H_0)$, and $P(p_\delta = 1|H_1)$, and prior probability ratio $r$, is instructive. Equation (4) explains how a study design influences the reliability of subsequent inference (SR19).

Every second-generation *p*-value inherits its reliability from the study design and the original odds of the alternative hypothesis. This reliability varies with the alternative hypothesis in question. As the distance between the alternative and null grows, the FDR and FCR decrease. However, the wide range of possible alternatives makes it hard to summarize the FDR and FCR with a single number. Summarization is even more of a problem in high-dimensions, such as our leukemia example, where every finding would ideally be accompanied by its estimated reliability.

Figure 5 displays the FDR and FCR (solid lines) when the width of the interval null is equal to one standard deviation ($\delta = \sigma/2$) and the sample size is large enough to permit $p_\delta = 1$. Included for comparison are the false discovery rate and false non-discovery rate of a comparable hypothesis test (dotted lines). These rates are $P(H_0|Rejected\ H_0) = [1 + r(1 - \beta)/\alpha]^{-1}$ and $P(H_1|Failed\ to\ Reject\ H_0) = [1 + (1 - \alpha)/\beta r]^{-1}$. The FDR and FCR for second-generation *p*- values are generally smaller than their hypothesis testing counterparts. SR20 displays this relationship for various sample sizes.

**Figure 5:** *Illustration of the false discovery rate (red) and false confirmation rate (blue) for second-generation p-values (solid lines). The false discovery rate (red) and false non-discovery rate (blue) from a comparable hypothesis test are shown as dotted lines. This example uses $r = 1, \alpha = 0.05, \delta = \sigma/2$, and $n = 16$, but the ordering of the curves is quite general.*

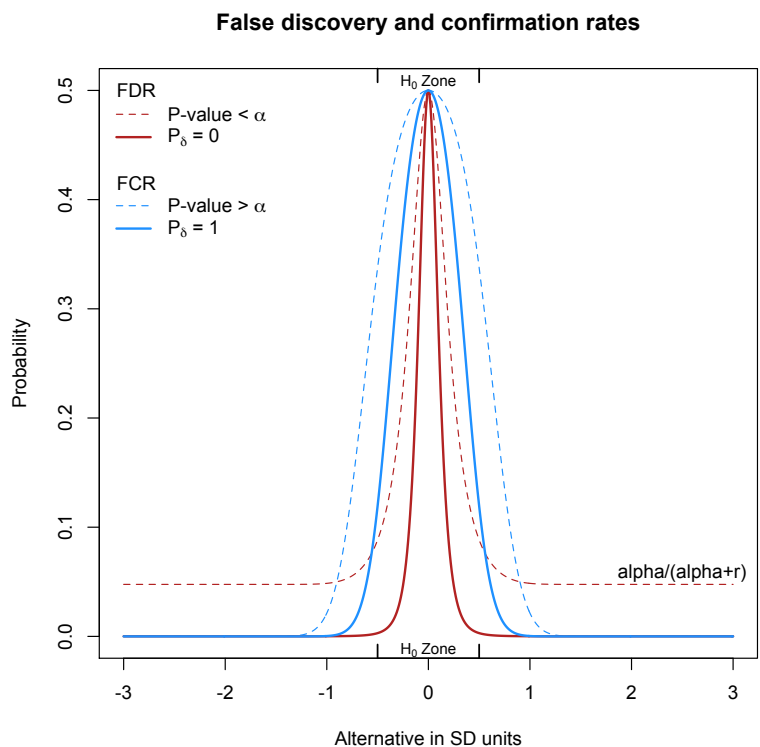

In principle, the reliability of any inferential summary ought to be reported along with the summary itself. Neither is truly sufficient on its own. However, reporting the reliability can be difficult to do in practice for the reasons noted above. And while standards for reporting the FDR and FCR deserve further investigation and discussion, the key message is that second-generation *p*-values decrease the FDR and FCR from classical *p*-values.

We note that while estimates further from the point null are more reliable in the sense that their false discovery rates are lower, this does not mean that they are more 'important', more 'relevant' or more 'meaningful' in a scientific sense. A low false discovery rate means only that the results are more likely to replicate under identical conditions. The cause for this replication could be a true scientific finding or an unknown experimental flaw. Only careful investigation into the validity of the scientific experiment can determine which it is.

## 5. Comments

More than a century of experience tells us that *p*-value usage, despite its flaws, will persist. Science desires an easily digestible and reliable summary of whether the data are compatible with null or alternative hypotheses. *Second-generation p-values are tailor made for this role*. They are easy to apply and interpret; they can be used in conjunction with frequentist, Bayesian or Likelihood methods; and they exhibit excellent frequency properties. Moreover, they eliminate the haggling over ad-hoc adjustments to *p*-values that have become a real challenge in high-dimensional and data-rich settings.

A well-known problem with classical *p*-values is that they can be small and yet also be associated with confidence intervals that include hypotheses or parameter values that are essentially null. In genomic studies with many thousands of variants, this has led to the detection of hundreds of variants with spurious significance. By introducing an interval null hypothesis of scientifically equivalent null hypotheses, second-generation *p*-values avoid the vast majority of the spuriously significant findings. By better reflecting the true nature of the null hypothesis in mathematical terms, statistical inference works better for science. Also, the width of the interval null need not be large, as the benefits of second-generation *p*-values will eventually be realized as the sample size grows.

Some challenges remain. The statistical properties of this new tool need to be explored and detailed in a variety of settings. Our preliminary findings detailed here, and in the web-supplement, indicate that second-generation *p*-values tend to have excellent behavior overall. Nonetheless, this should be explored in detail. In practice, disagreement on the width of the interval null ($\delta$) will require re-calculation of the second-generation *p*-value. To facilitate this, we encourage the reporting and discussion of the interval estimates upon which the second-generation *p*-value is based. We remind the reader that the second-generation *p*-value is not intended to be the final product, but rather a quickly digested summary measure. Understanding the force and implication of statistical results will always require attention to the details of the statistical analysis.

Regardless of the analysis, caution should always be used when interpreting statistical findings. Just because a result is statistically less likely to be a 'false discovery' does not mean that the result is more 'important', 'relevant' or 'meaningful' in a scientific sense. It means simply that the results are more reliable and more likely to replicate under identical circumstances, even if those circumstances are in some way flawed. A statistical hypothesis is a specific, precise mathematical statement about an unknown parameter. And statistical hypotheses assume a probability model to perform computations. If that model fails, then the associated statistical hypotheses often fail to be valid translations of the scientific hypotheses. This is, in fact, one of the most common criticisms of point null hypothesis testing; the point null is almost never *exactly* correct (employing an interval null hypothesis addresses this concern directly). In this paper, we assumed that the broader statistical model holds because this is a routine assumption for *p*-value based inference. However, the robustness of the second-generation *p*-value to model misspecification is an important topic that deserves attention.

We anticipate that the major challenge in implementing second-generation *p*-values will be encouraging researchers to define a scientifically relevant finding *prior to examining the data*. While this is already routine in some areas of biomedicine (e.g. clinical trials), it is not even on the radar in others. Nevertheless, second-generation *p*-values are a clear improvement over classical *p*-values. They are between 0 and 1. They have a straightforward interpretation. A second-generation *p*-value of 0 is properly interpreted as favoring alternative hypotheses. A second-generation *p*-value of 1 is properly interpreted as favoring null hypotheses. A second-generation *p*-value between 0 and 1 favors both types of hypotheses and is properly interpreted as inconclusive. The error rates for second-generation *p*-values are bounded by $\alpha$ and converge to 0. Adjustments for multiple comparisons are obviated, and lower false discovery rates accompany observed results. In short, *the second-generation p-value achieves the inferential properties that many scientists hope, or believe, are attributes of the classic p-value.* Using second-generation *p*-values can only improve rigor, reproducibility and transparency across science.

## Author Contributions

Second-generation *p*-values were first conceptualized and defined by JDB. All authors contributed to the refinement of the second-generation *p*-value concept, the investigation of its frequency properties, the development and coding of examples presented in this paper, and the drafting of this manuscript.

## Acknowledgements

The authors acknowledge Valerie Welty, Karen Bandeen-Roche, David Colquhoun, two anonymous referees, and the academic editor, Neil R. Smalheiser, for critical comments that improved the manuscript.

## Supplementary Materials:

Figures S1-S6

References (27-32)

**Data Sharing Statement:**
The Golub data are publicly available online via the 'golubEsets' package in Bioconductor. They can also be accessed at https://github.com/ramhiser/datamicroarray/wiki/Golub-(1999) or http://portals.broadinstitute.org/cgi-bin/cancer/publications/pub_paper.cgi?mode=view&paper_id=43.

**Supplementary Materials**

**Title:** **Second-generation** *p*-values: improved rigor, reproducibility & transparency in statistical analyses

**Authors:** Jeffrey D. Blume[1*] PhD, Lucy D'Agostino McGowan[2] MS, William D. Dupont[3] PhD, Robert A. Greevy[1] Jr. PhD

**Affiliations:**
[1]Associate Professor, Department of Biostatistics, Vanderbilt University School of Medicine, Nashville TN
[2]PhD Candidate, Department of Biostatistics, Vanderbilt University School of Medicine, Nashville TN
[3]Professor, Department of Biostatistics, Vanderbilt University School of Medicine, Nashville TN

[*]Correspondence to: j.blume@vanderbilt.edu

**Contents:**

**Frequently Asked Questions (FAQ) about second-generation *p-value*s**

Q.1:     Why is it called a 'second-generation' *p*-value?

Q.2:    How does one interpret a second-generation *p*-value?

Q.3:    Is the 'second-generation' *p*-value a proportion or probability?

Q.4:    Is the second-generation *p-value* the posterior probability of the null hypothesis assuming a non-informative prior?

Q.5:    What is the second-generation *p-value* estimating?

Q.6:    Why can't I interpret a traditional *p*-value in the way that a second-generation *p*-value is interpreted?

Q.7:    Why is the second-generation *p-value* more useful than the simple report of the 95% confidence interval?

Q.8:    Is it a problem that second-generation *p-values* of zero can be associated with different levels of precision?

Q.9:    Do I need to use the same interval null if I want to compare second-generation *p-values*?

Q.10:   Why do we need three regions for interpreting data?

Q.11:   Why do we need to consider an interval null hypothesis?

Q.12:   Why are very small differences, say near zero changes, between populations not scientifically meaningful?

Q.13:   Who determines the width of the null interval?

Q.14:   Why is the null interval uniform around the point null and not some other shape?

Q.15:   How do we guarantee that the null interval would be defined before data were collected?

Q.16:   Why have we been using point null hypotheses for so long?

Q.17:   Can a classical p-value be computed for an interval null hypothesis?

Q.18:   Why do I have to set the interval null before looking at the data?

Q.19:   Can I report the smallest null interval for which the second-generation *p-value* is still zero?

Q.20:   What is wrong with assessing scientific meaningfulness after the analysis is complete?

Q.21:   Won't second-generation *p-values* be harder for non-statisticians to understand?

Q.22:   Isn't the problem with traditional *p*-values that we have to choose an arbitrary cutoff? Can't this be fixed by finding the right cut-off?

Q.23:   Is the precision of a single data point really relevant in determining the interval null width?

Q.24:   We don't know much about second-generation *p-values*, so should we use it?

**Q.1:  Why is it called a 'second-generation' *p*-value?**

Ans:  Our proposed new metric is a conceptual generalization of the *p*-value that is rooted in the duality of confidence intervals and hypothesis testing. Hence the "second-generation" tag seems appropriate. Instead of checking to see if a singular null hypothesis is the interval, we now check to see how many of the practical representations of the null hypothesis are in the interval. We report how many of the best supported hypotheses are null hypotheses (e.g., all of them, $p_\delta = 1$, or none of them, $p_\delta = 0$).

**Q.2:  How does one interpret a second-generation *p*-value?**

Ans:  Second-generation *p*-values have a very natural interpretation: the fraction or proportion of data-supported hypotheses that are null hypotheses. Second-generation *p*-values are descriptive statistics that are intended to summarize the degree to which the study generated support for the null hypotheses or alternative hypotheses. A $p_\delta = 1$ or $p_\delta = 0$ indicates that the study has reached a natural stopping point.

**Q.3:  Is the 'second-generation' *p*-value a proportion or probability?**

Ans:  Second-generation *p*-values are proportions. They are not estimates of some unknown population quantity; they are not an estimate of the probability that the null hypothesis is true (neither is a first-generation *p*-value, for that matter). The *p* in $p_\delta$ is intended to stand for proportion not probability. It can be helpful to think of second generation *p*-values as an indicator of when the study has reached its goal.

**Q.4:  Is the second-generation *p*-value the posterior probability of the null hypothesis assuming a non-informative prior?**

Ans:  No. The posterior probability is P( Null | Data), which is not a second-generation *p*-value if only because the conditioning set is different.

**Q.5:** **What is the second-generation *p*-value estimating?**

Ans: Nothing. It is a descriptive statistic. It describes the proportion of the data-supported hypotheses that are null hypotheses. In large samples, the second-generation *p*-value converges to zero or one, and is thus best thought of as a marker of when an experiment reaches its predetermined goal.

**Q.6:** **Why can't I interpret a traditional *p*-value in the way that a second-generation *p*-value is interpreted?**

Ans: Because the theory of significance testing is quite clear: traditional *p*-values have a uniform distribution under the null hypothesis. Hence, the magnitude of non-significant *p*-values is indicative of nothing more than randomness; any number larger than $\alpha$ is inconclusive. It is impossible, by design, for a traditional *p*-value to represent evidence for the null hypothesis. This is, in part, why replicating and interpreting non-significant *p*-values is so problematic. Under the null hypothesis *p*-values are just random variables that bounce around between 0 and 1.

**Q.7:** **Why is the second-generation *p*-value more useful than the simple report of the 95% confidence interval?**

Ans: It is more useful for providing a quick assessment of whether the experiment met is pre-determined goals. It saves time and is easier to comprehend than assessing the overlap between a 95% CI and unstated hypotheses of scientific interest. Second-generation *p*-values are not a replacement for 95% CIs. Confidence intervals are important in that they provide information on the range of effects and level of precision supported by the data.

**Q.8:** **Is it a problem that second-generation *p*-values of zero can be associated with different levels of precision?**

Ans: No. The second-generation *p*-value is not meant to replace the confidence interval, but rather to augment it. It is only seeking to convey if any null hypotheses were among those best supported by the data. The data can be imprecise and still exclude all null hypotheses. When comparing two second-generation *p*-values of zero, we recommend comparing their delta-gaps (the distance from interval null to interval estimate in SD units).

**Q.9: Do I need to use the same interval null if I want to compare second-generation *p*-values?**

Ans: Yes, this is needed as well as the same type of interval estimate if you want to compare magnitudes of second-generation *p*-values. Note this is similar to traditional *p*-values, which are not comparable unless they are based on the same sample size. When the second-generation *p*-value is zero, we recommend using the delta gap – the distance from interval null to interval estimate in $\delta$ units – as a way to rank second generation *p*-values that are zero.

**Q.10: Why do we need three regions for interpreting data?**

Ans: The most important thing we can do is not misrepresent inconclusive results. Ideally, every experiment would conclude with data that are only compatible with either the null hypothesis ($p_\delta = 1$) or alternative hypothesis ($p_\delta = 0$). But with finite sample sizes, this is often not possible. This is a major issue for classical *p*-values that is resolved by second-generation *p*-values.

**Q.11: Why do we need to consider an interval null hypothesis?**

Ans: Point null hypotheses are neither detectable nor a practical reality. Measurement devices have limited precision and many small non-zero changes are not actionable, consequential, or reproducible. As a result, for any experiment, there is actually a range of near null effects that are indistinguishable from the point null hypothesis and inferentially inconsequential. These are all null hypotheses and there is no reason to discriminate between them.

**Q.12: Why are very small differences, say near zero changes, between populations not scientifically meaningful?**

Ans: This depends on context. If the difference can be measured on an individual unit, then it can be meaningful and lead to action or intervention. Often, we find differences between populations are within the precision of the instrument. In that case, it is not clear if the difference is "real" or due to measurement error. For example, if the average income between two states were found to differ by one-half of one cent, would this be meaningful or actionable? What about one-tenth of one cent? There is always some point at which the measurement scale effectively becomes discrete.

**Q.13: Who determines the width of the null interval?**

Ans: The researcher sets this benchmark when designing the study. This is often implicitly done when power projections are completed. In applications such as clinical trials, this often is justified and discussed. But the researcher sets his or her own benchmark.

**Q.14: Why is the null interval uniform around the point null and not some other shape?**

Ans: The interval null is neither uniform nor any other shape. No distributional assumptions about the interval null are needed to use a second-generation *p*-value. Simplicity is the motivating factor here.

**Q.15: How do we guarantee that the null interval would be defined before data were collected?**

Ans: We can't. It is always possible to cheat. It is just as easy to cheat with *p*-values; it is called *p*-hacking. Protections such as pre-specified analysis plans will work similarly for second-generation p-values as they do for first generation p-values.

**Q.16: Why have we been using point null hypotheses for so long?**

Ans: Allowing the statistical procedure to pretend it has infinite precision is a welcome mathematical convenience. But the price of convenience is high, e.g. confusion as to whether statistical significance imparts a consequential finding. And when the assessment of clinical or scientific significance is forgotten, ignored, or determined after looking at the data, the results are deemed suspect. This last concern is real; despite our best efforts, the definition of a meaningful effect tends to change after the data are observed.

**Q.17: Can a classical *p*-value be computed for an interval null hypothesis?**

Ans: Classical *p*-values for interval null hypotheses are undefined. This is because there exists a set of *p*-values, one for every null hypothesis. No single number summary of the *p*-value set captures, in general, how compatible the data are with the interval null hypothesis. It might be that the maximum *p*-value best tells the story in one case, while in another the minimum *p*-value or a weighted average of *p*-values is best. The reason for this is that the data may support some parts of the null set and not others.

**Q.18: Why do I have to set the interval null before looking at the data?**

Ans: This is good experimental practice: defining what success means before the experiment is conducted. Otherwise, the data are used twice: once to set the interval null hypothesis and once to check if the data are compatible with the interval null. This dual use of the data prevents the analysis from being confirmatory, increases the chances that the results are a false positive, and reduces the chances that the results can be reproduced.

**Q.19: Can I report the smallest null interval for which the second-generation *p*-value is still zero?**

Ans: Sure, but such an analysis would be considered exploratory. Instead, we suggest reporting the delta-gap (the distance from interval null to interval estimate in SD units) when the second-generation *p*-value is zero.

**Q.20:  What is wrong with assessing scientific meaningfulness after the analysis is complete?**

Ans:  The problem with checking scientific relevance after establishing statistical significance is that it allows after-the-fact rationalization, which reduces the rigor and reproducibility of published results. Requiring that the smallest effect of interest be specified up front, as is done with second-generation *p*-values, improves scientific rigor and reproducibility. In many applications, such as clinical trials, the interval null is already implicitly specified upfront for power projections.

**Q.21:  Won't second-generation *p*-values be harder for non-statisticians to understand?**

Ans:  Not in our experience. Non-statisticians readily acknowledge that there is always a null region of effects, and they readily accept an interval of hypotheses that are supported by the data. These are routine elements of current practice. The intersection of these two intervals follows naturally and is easy to conceptualize. Some understanding of confidence intervals and hypothesis testing is necessary for power projections or for understanding their frequency properties. But an advanced level of understanding is not required for interpreting observed data with second-generation *p*-values.

**Q.22: Isn't the problem with traditional *p*-values that we have to choose an arbitrary cutoff? Can't this be fixed by finding the right cut-off?**

Ans:  No. The problem is the metric itself. Choosing a different cutoff would not resolve issues with interpretation, computation or reproducibility. This is because adjustments like Bonferroni search for findings in *p*-value space where scientific meaningfulness is obfuscated. In contrast, second-generation *p*-values search for findings only in the space of scientifically meaningful results. Moreover, second generation *p*-values classify results into three categories (supporting the null hypothesis, inconclusive, supporting the alternative hypotheses), while traditional *p*-values use only two (inconclusive and supporting the alternative hypotheses).

**Q.23:** **Is the precision of a single data point really relevant in determining the interval null width?**

Ans: That is just one method of determining the width of the interval null hypothesis. It can be set in many ways. However, it makes little sense to search for differences between populations that cannot be measured in an individual. For example, what would it mean if a drug were found to reduce systolic blood pressure by 1 mmHG? If we gave that drug to an individual, the reduction in blood pressure would not be measurable nor is it considered clinical meaningful. Moreover, it would not be clear if the difference is due to measurement error or a real effect.

**Q.24:** **We don't know much about second-generation *p*-values, so should we use it?**

Ans: Yes. While there is certainly more to explore, this should not reduce enthusiasm for potentially unifying advance. Besides, second-generation *p*-values are essentially a formalization of current practice that is more likely to be reproducible and transparent.

**Main Supplementary Text:** Remarks are referenced as 'Supplement Remark #' or (SR#).

*Remark 1.* Naturally, neither $I$ nor $H_0$ may contain the entire parameter space. Other pathologies are easily rectified. For example, if intervals $I$ and $H_0$ overlap, but $I \subseteq H_0$, i.e., $I$ is a subset of $H_0$, then $p_\delta = 1$ regardless of the length of the intervals. The problem arises when the intersection is finite, $|I \cap H_0| < \infty$, but both intervals are not. For example we might have $I = [c, \infty)$ and $H_0 = (-\infty, d]$ with $c < d$ real numbers. Now we have $|I \cap H_0| = d - c$ and we could argue that $|I|/|H_0| = 1$. But $|I \cap H_0|/|I|$ is arguably zero, whereas $p_\delta = 0$ seems inappropriate here because the intervals have a finite set of hypothesis in common. A practical and realistic solution is to simply truncate $I$ at effects that are not possible to observe in practice.

*Remark 2.* Note that the procedure is inferentially consistent for all null and alternative hypothesis that are not on the boundary of the indifference zone. When the true hypothesis is exactly on the boundary of the interval null, say at $\mu_0 + \delta$, the second-generation *p*-value will have essentially the same frequency properties as a classical hypothesis test. As a result, the Type I Error rate of $\alpha$ will remain constant as a function of the sample size and the procedure is no longer inferentially consistent in the limit. That is, it will be wrong $100\alpha\%$ of the time regardless of the sample size.

*Remark 3.* An underappreciated fact of statistical inference is that over 99% of Type I Errors occur between 1.96 to 4 *standard errors* from the null. Because $2 * (\Phi[4] - \Phi[1.96])/0.05 = 0.9986$ where $\Phi[x] = P(Z \leq x)$ is the standard normal cumulative distribution function. In moderate to large samples, alternative hypotheses in this region are very close to the null hypothesis in absolute units and seldom represent a practically different model than the null. Because $p_\delta$ excludes alternative hypothesis that are scientifically indistinguishable from the point null hypothesis – even if statistically significant – the rate of Type I Errors is dramatically reduced.

*Remark 4.* The exact relationship will depend on circumstances, but this simple case provides a good guide. Let $E$ be the margin of error (half width) from a $(1-\alpha)100\%$ CI with sample size $n$. Also, let $n$ be the sample size from a two-sided hypothesis test with size $\alpha$ and power $1-\beta$ to detect an alternative that is $\delta$ units from the null hypothesis. Assuming the variance is constant, we have that

$$n = \left(\frac{Z_{1-\alpha/2} + Z_{1-\beta}}{\delta}\right)^2 = \left(\frac{Z_{1-\alpha/2}}{E}\right)^2 \tag{S1}$$

If we let $\delta$ represent the smallest change of scientific interest, then $2\delta = |H_0|$ and $|I| = 2E$ and we have that

$$|I| = \left(\frac{Z_{1-\alpha/2}}{Z_{1-\alpha/2} + Z_{1-\beta}}\right)|H_0| \tag{S2}$$

where $Z_{1-\alpha/2} = \Phi^{-1}[1-\alpha/2]$ and $\Phi[x] = P(Z \le x)$ is the standard normal cumulative distribution function. We find that $|I| = |H_0|$ when the sample size confers 50% power to detect $\delta$. With 80% and 90% power, we have $|I| = 0.7|H_0|$ and $|I| = 0.6|H_0|$, respectively, when $\alpha = 0.05$. It follows that the power has to drop below 16% for the correction factor to be triggered.

*Remark 5.* In future work, we intend to use 1/8 likelihood support intervals for the basis of our second-generation *p*-values. This is easily achieved with standard software by using a 96% CI from a normal approximation when the underlying sampling distribution is symmetric (which is most settings) [21].

*Remark 6.* It is instructive to see what happens as the indifference zone for fold change shrinks. Only 264 genes have Bonferroni corrected *p*-values less than 0.05, while 1233 genes have empirical false discovery rates (q-values) less than 0.05. There are 2028 genes with raw *p*-values less than 0.05. The ordered raw *p*-values are displayed in black along with Bonferroni adjusted *p*-values red and FDR/q-values (blue). Here $\alpha = 0.05$ in all cases. This is depicted in Supplementary Figure 1S. The indifference zones are $\delta = 0.3, 0.1, 0.05, 1 \times 10^{-6}$. The second-generation *p*-values converge to $I(P > 0.05)/2$. The figure makes it clear second-generation *p*-values are doing something different than a routine multiple comparisons adjustment. Standard adjustments are made in *p*-value space where the effect size and variance are confounded. In contrast, second generation *p*-values are effectively making adjustments based on the observed effect size and screening out the results that are more likely to be false discoveries, i.e. the significant effects that are also null or practically null effects.

*Remark 7.* Figure S2 shows how the second-generation *p*-values were computed to color the rug plot. The estimated survival differences are plotted with their confidence interval and the indifference zone (shaded region). The confidence interval on the difference in survival rates could be computed using asymptotic methods or a simple bootstrap. Here we used the variance of the predictions from a cox proportional hazard model and assumed the two groups were independent. An alternative approach would be to estimate the baseline hazard using some other non-parametric method.

*Remark 8.* The 2x2 table examines a binary exposure's association, say smoking, with a binary outcome, say lung cancer. Imagine 100 smokers and 100 non-smokers, where 65 smokers and 50 non-smokers developed lung cancer. This is displayed in Table S1.

**Table S1: Mock outcomes from a cohort study**

| Exposure | Outcome | |
|---|---|---|
| | Lung CA | No Lung CA |
| Smoker | 65 | 35 |
| Non-Smoker | 50 | 50 |

The odds ratio of 1.86 measures the association between smoking and cancer. Here we have,

$$or = \frac{P(cancer|smoker)P(nocancer|nonsmoker)}{P(cancer|nonsmoker)P(nocancer|smoker)} = \frac{65 \times 50}{35 \times 50} = 1.86$$

Statistical computations are usually done on the natural logarithm scale, yielding a log odds ratio of 0.62 and a 95% CI of 0.05 to 1.19. The traditional *p*-value of 0.032 indicates statistical significance, rejecting the null hypothesis that the log odds ratio is 0. However, with an interval null of -0.1 to 0.1 (a 10% change in the odds ratio), we have $p_\delta = 0.044 = \frac{(0.1-0.05)}{(1.19-0.05)}$ (1) indicating the data are inconclusive and only suggestive of a real effect. The magnitude of an inconclusive second-generation *p*-value can vary slightly when the effect size scale is transformed. However definitive findings, i.e. a $p_\delta$ of 0 or 1, are *not* affected by the scale changes.

Like traditional *p*-values, the scale of the effect size can matter for reporting findings. Consider the contingency table example in section 3.1. Using the natural odds ratio scale (anti-log), the 95% CI is 1.05 to 3.29. The second-generation *p*-value is now slightly less at $p_\delta = 0.024 = \frac{(1.11-1.05)}{(3.29-1.05)}$ (1). While the conclusion is essentially the same, the degree to which the data are deemed "inconclusive" varied slightly. Importantly, a $p_\delta$ of 0 or 1 will *not* be affected by monotonic scale changes (virtually all inferential transformations are monotonic). When a second-generation *p*-value indicates complete (in)compatibility with the null hypothesis, the results are invariant to the analysis scale.

*Remark 9.* Data on glycohemoglobin from the 2009-2010 National Health and Nutrition Examination Survey (NHANES), was designed to assess the health and nutritional status of adults and children in the United States (27,28). Hemoglobin A1c (HbA1c), is a measure of the amount of glucose bound to hemoglobin in red blood cells and is a popular biomarker in diabetes and cardiovascular disease research.

Suppose we have the following linear regression model for HbA1c

$$HbA1c = \beta_0 + \beta_1 age + \beta_2 sex + \beta_3 race$$
$$+\beta_4 weight + \beta_5 waist + \beta_6 triceps + \epsilon$$

(S3)

with independent errors $\epsilon \sim N(0, \sigma^2)$. Taken together, weight, waist size, and triceps thickness represent the impact of body size on HbA1c. We can assess the contribution of body size to this model and remove these predictors if they do not contribute sufficiently. This is usually posed as a problem of determining if the parameter vector $[\beta_4, \beta_5, \beta_6]$ is sufficiently close to $[0,0,0]$.

We could compute the second-generation *p*-value for the three-dimensional vector, but this requires specifying a three-dimensional interval null and obtaining simultaneous confidence intervals, i.e. a CI for the entire vector as opposed to three independent CIs for each element. A more elegant approach is examining how much the explained variance decreases when body size is removed from the model. That is, how different are the coefficients of determination ($R^2$) between the full and reduced model, say $R_f^2 - R_r^2$. Routine alternatives are described below. A CI for this difference is easily bootstrapped. Algina et al. (30) advocate for the delta method approximation by Alf (31), while Smithson (32) uses the non-central F-distribution.

The three 95% CIs were 0.0231 to 0.0427 (BCa bootstrap), 0.0251 to 0.04107 (delta method), and 0.0246 to 0.0405 (non-central F). Suppose our null interval for describing an impactful contribution is $H_0: 0 \leq R_f^2 - R_r^2 \leq 0.025$. The resulting second-generation *p*-values are 0.097, 0, and 0.024. Notice that the choice of confidence interval method turns out to be important, so this should be carefully considered before examining the data. The two conservative methods indicate the data are only suggestive of a contribution and still inconclusive. If, however, we consider a more stringent criterion, such as $H_0: 0 \leq R_f^2 - R_r^2 \leq 0.05$, all three second-generation *p*-values would be 1, indicating the improvement in $R^2$ is less than 5%.

It is sometimes helpful to benchmark the reduction in explained variation against the total amount of unexplained variation that results when only the reduced model is fit. This has a direct tie to the tradition partial f-test or 'chunk-test'. This can be obtained by simply re-scaling the obtained CIs by $1 - R_r^2$ so that the estimand of interest is $R_f^2 - R_r^2 / (1 - R_r^2)$, which is just the

 

squared partial correlation $R_{f.r}^2$. The routine ANOVA F-statistic for this comparison is just

$R_{f.r}^2(df_f)/(1 - R_{f.r}^2)(df_f - df_r)$. See Smithson (32) for details.

*Remark 10.* **Statistical and frequency properties of second-generation *p*-values**

There are three cases to consider: the probability data are compatible with the alternative, $P(p_\delta = 0)$, the probability data are compatible with the null, $P(p_\delta = 1)$, and the probability data are inconclusive $P(0 < p_\delta < 1)$. Note that we have three potential outcomes to consider instead of just two ("Reject the null" or "Fail to reject the null"). In Remarks 11 through 18, we examine the statistical properties of second-generation *p*-values when the sampling distribution of the estimator can be approximated by a normal distribution. This scenario covers a large majority of statistical applications, including methods of moments and maximum likelihood estimation, as well as common non-parametric estimators in large samples.

*Remark 11.* **Distributional assumptions:** Let $\hat{\theta}_n$ be an estimator of parameter $\theta$. We consider the case where the sampling distribution is $\sqrt{n}(\hat{\theta}_n - \theta) \overset{A}{\sim} N(0, V)$ where the variance $V$ is known or can be readily estimated. This scenario reflects the core behavior of a large majority of statistical applications, such as methods of moments, maximum likelihood estimation, and some common non-parametric estimators in large samples, i.e., U-statistics.

*Remark 12.* **Observing data compatible with the alternative hypothesis:** How often will a given set of data indicate compatibility with the alternative hypothesis? This probability, $P(p_\delta = 0)$, is analogous to power. Since $p_\delta$ is 0 only when the intersection between the intervals is the empty set, it follows that

$$P_\theta(p_\delta = 0) = \Phi\left[\frac{\sqrt{n}(\theta_0 - \delta)}{\sqrt{V}} - \frac{\sqrt{n}\theta}{\sqrt{V}} - Z_{\alpha/2}\right] + \Phi\left[-\frac{\sqrt{n}(\theta_0 + \delta)}{\sqrt{V}} + \frac{\sqrt{n}\theta}{\sqrt{V}} - Z_{\alpha/2}\right] \quad \text{(S4)}$$

where $\theta_0$ is the point null hypothesis and $\theta$ is the 'true' data generating hypothesis. As expected, the 'power curve' is a function of $\delta$, the indifference zone margin.

When graphed, it looks like a power curve that was cut in half and pulled apart. Figure S3 displays the power for an interval null hypotheses of the form $H_0: \theta_0 - \delta \le \theta \le \theta_0 + \delta$, which was graphed for $\delta = 0, 0.03, 0.5, 1$. The zero origin on the x-axis represents $\theta_0$. The loss in power is understandable. It is no longer sufficient to just rule out $\theta_0$, we must rule out all of the null hypotheses. This effectively changes the anticipated effect size from $\theta_{alt} - \theta_0$ to $\theta_{alt} - \theta_0 - \delta$ which accounts for the power loss. Alternatives in the indifference zone are null hypotheses and the curve over that section of the x-axis represents the usual Type I Error rate. It should be clear from the graph that the maximum Type I error rate for second-generation *p*-values is bounded by $\alpha$ at all sample sizes (SR13). Type I Error rates near $\alpha$ occur when the true hypothesis is near or on the edge of the indifference zone.

*Remark 13.* When $\theta = \theta_0$, Equation S4 is analogous to the Type 1 Error rate which reduces to

$$P_{\theta_0}(p_\delta = 0) = 2\Phi\left[-\frac{\sqrt{n}\delta}{\sqrt{V}} - Z_{\alpha/2}\right] \quad \text{(S5)}$$

Note the dependence on the sample size $n$ and $\delta$. Hence, the Type I Error rate is bounded above by $2\Phi\left[-Z_{\alpha/2}\right] = \alpha$. Moreover, it shrinks to 0 as the sample size approaches infinity, for any given $\delta > 0$. When $\delta = 0$, we recover the usual Type I Error rate.

*Remark 14.* **Comparison with a Bonferroni adjustment:** Figure S4 displays $P(p_\delta = 0)$ using an indifference zone of -0.3 to 0.3 similar to that from the Leukemia microarray example. For comparison, we added the power curves from a single classical hypothesis test and from a Bonferroni correction procedure with 10, 1000, and 7128 comparisons. With 10 comparisons and a fixed variance, the Bonferroni and second-generation *p*-values have virtually identical operating characteristics (See SR16 for discussion). As seen in the microarray example, the second-generation *p*-value outperforms the Bonferroni adjustment with a large number of heterogeneous comparisons. Second-generation *p*-values achieve the benefit of traditional multiple comparisons adjustments through the use of a scientific adjustment instead of an ad-hoc statistical adjustment.

*Remark 15.* Just because two statistical procedures have nearly identical frequency properties does not imply that they will yield the same findings. A case in point is the second-generation *p*-value and Bonferroni procedure with k=10 comparisons in our example. Their power curves are nearly the same (Figure 6), implying they will have similar Type I and II error rates and similar family wise error rates. However, as the Leukemia example demonstrates, these procedures result in very different findings. Hence the Bonferroni procedure with k=10 is no substitute for a second-generation *p*-value.

*Remark 16.* **Observing data compatible with the null:** An important advance of second-generation *p*-values is that they can indicate when the data are compatible with the null hypothesis. How often does this happen? The data are compatible with the null hypothesis when the interval null contains the entire interval estimate. When the width of the interval estimate is less than the width of the interval null hypothesis, i.e., $|I| < |H_0|$ or $\delta > Z_{\alpha/2}\sqrt{V/n}$, we have

$$P_\theta(p_\delta = 1) = \Phi\left[\frac{\sqrt{n}(\theta_0 + \delta)}{\sqrt{V}} - \frac{\sqrt{n}\theta}{\sqrt{V}} - Z_{\alpha/2}\right] - \Phi\left[\frac{\sqrt{n}(\theta_0 - \delta)}{\sqrt{V}} - \frac{\sqrt{n}\theta}{\sqrt{V}} + Z_{\alpha/2}\right] \tag{S6}$$

Otherwise, $P_\theta(p_\delta = 1) = 0$. Also, when the point null is the 'true' hypothesis, i.e., $\theta = \theta_0$, (S6) reduces to

$$P_{\theta_0}(p_\delta = 1) = \begin{cases} \Phi\left[\frac{\sqrt{n}\delta}{\sqrt{V}} - Z_{\alpha/2}\right] - \Phi\left[-\frac{\sqrt{n}\delta}{\sqrt{V}} + Z_{\alpha/2}\right] & \text{for } \delta > Z_{\alpha/2}\sqrt{V/n} \\ 0 & \text{o. w.} \end{cases} \tag{S7}$$

Notice that when $\delta = Z_{\alpha/2}\sqrt{V/n}$, this expression is 0 because $P_{\theta_0}(\hat{\theta}_n = \theta_0) = 0$ by definition. Figure S5 displays $P(p_\delta = 1)$ as a function of the indifference zone margin $\delta$ when the sample size is small (right panel) and when the sample size is large (left panel). For very small indifference zones, it is virtually impossible to observe enough data to demonstrate compatibility with the null hypothesis. For large indifference zones, the data indicate compatibility with the null hypothesis most often when the true hypothesis is near the middle of the interval null.

*Remark 17.* **Observing data that are inconclusive:** Perhaps the scourge of any study is inconclusive results. Here we detail the probability that the second-generation *p*-value is inconclusive. The probability of observing data that are inconclusive is:

$$P_\theta(0 < p_\delta < 1) = 1 - \Phi\left[\frac{\sqrt{n}(\theta_0 - \delta)}{\sqrt{V}} - \frac{\sqrt{n}\theta}{\sqrt{V}} - Z_{\alpha/2}\right] - \Phi\left[-\frac{\sqrt{n}(\theta_0 + \delta)}{\sqrt{V}} + \frac{\sqrt{n}\theta}{\sqrt{V}} - Z_{\alpha/2}\right]$$

$$- \Phi\left[\frac{\sqrt{n}(\theta_0 + \delta)}{\sqrt{V}} - \frac{\sqrt{n}\theta}{\sqrt{V}} - Z_{\alpha/2}\right] + \Phi\left[\frac{\sqrt{n}(\theta_0 - \delta)}{\sqrt{V}} - \frac{\sqrt{n}\theta}{\sqrt{V}} + Z_{\alpha/2}\right]$$

(S8)

when $\delta > Z_{\alpha/2}\sqrt{V}/\sqrt{n}$ and

$$P_\theta(0 < p_\delta < 1) = 1 - \Phi\left[\frac{\sqrt{n}(\theta_0 - \delta)}{\sqrt{V}} - \frac{\sqrt{n}\theta}{\sqrt{V}} - Z_{\alpha/2}\right] - \Phi\left[-\frac{\sqrt{n}(\theta_0 + \delta)}{\sqrt{V}} + \frac{\sqrt{n}\theta}{\sqrt{V}} - Z_{\alpha/2}\right]$$

(S9)

otherwise. Figure S6 displays this behavior. When the indifference zone is small relative to the intended precision, and the true hypothesis is in or near the indifference zone, the probability of inconclusive results is high. As the indifference zone widens, the probability of inconclusive results remains high at its edges when the truth is also at the edges. But the probability drops rapidly near the middle of the zone when such results would indicate compatibility with the null hypothesis. The take home message here is that data will tend to be inconclusive when the truth is near the edges of the indifference zone. The practical solution to this problem is to use an indifference zone that is neither too large nor too small. But, of course, this is much easier said than done.

*Remark 18.* The delta-gaps are: Here $\delta = \log_{10} 2 = 0.3$ and the null interval is $[-0.3, 0.3]$. Gene #2288 has a 95% CI of 2.11 to 2.87 on the $\log_{10}$ scale with a delta gap of $6.03 = (2.11 - 0.3)/0.3$. Gene #3252 has a 95% CI of 1.22 to 1.64 with a delta gap of $3.07 = (1.22 - 0.3)/0.3$.

*Remark 19.* Both the false discovery rate (FDR), $P(p_\delta = 0|H_0)$, and the false confirmation rate (FCR), $P(p_\delta = 1|H_1)$, in (4) converge to zero as the sample grows because the probabilities, $P(p_\delta = 0|H_1)$ and $P(p_\delta = 1|H_0)$, converge to zero. This does not happen for the FDR in hypothesis testing, which converges to $\alpha/(\alpha + r)$, because the Type I Error rate is held constant at $\alpha$. So design choices in this regard are consequential.

*Remark 20.* In the example used in the paper the FDR and FCR for second-generation *p*-values are smaller than their hypothesis testing counterparts. This is generally true for the FDR when the multiple comparisons being made have varying standard errors. However, it is possible for the FCR to be larger than the false non-discovery rate. This happens for hypotheses inside the null interval when the sample size is very large. As such is it not of consequence. By design, hypotheses within the indifference zone are not detectable by second-generation *p*-values. We believe this can be addressed by allowing the null interval to shrink at a rate slower than the interval estimate, but this will be detailed elsewhere. Figure S7 displays the FDR and FCR as the sample size changes. Note that the FCR is undefined in the first plot because the sample size is too small to permit nesting of the interval estimate in the interval null hypothesis.

*Remark 21.* In the pathological case when the variance across the multiple comparisons is constant, it is possible to select a $\delta$ such that the second-generation *p*-value's FDR would match the FDR of a Bonferroni correction. This connection highlights how the second-generation *p*-value outperforms the Bonferroni in a general setting. To have the same FDR as the Bonferroni, it would need to use a $\delta$ that depends on the standard error of each comparison. This is precisely what the second-generation *p*-value avoids, allowing it to detect large effects with relatively larger standard errors and while avoiding clinically meaningless effects that have relatively small standard errors.

**Figure S1:** *First and second-generation p-values plotted versus their ranking (number of rejected hypotheses). The black line gives unadjusted p-values, the blue line gives p-values after FDR adjustment, and the red line gives p-values after Bonferroni adjustment. Here $\alpha = 0.05$ in all cases. The green points are the second-generation p-values that result from the indifference zone in the title. Note the indifference zone narrows with $\delta = 0.3, 0.1, 0.05,$ and $1 \times 10^{-6}$.*

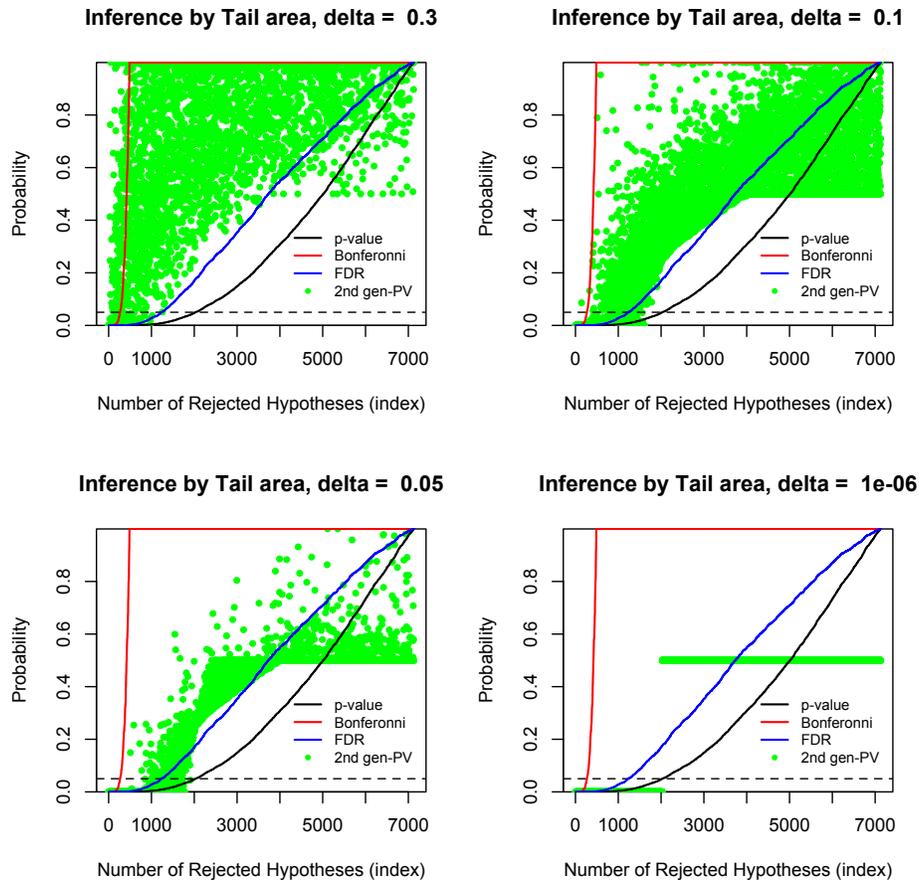

**Figure S2:** *The difference in survival fraction (black line) for in patients with advanced lung cancer from the North Central Cancer Treatment Group study. Red dashed lines are 95% confidence intervals. The indifference zone of +/- 5% is plotted in blue-grey. Rug plot on x-axis displays second generation p-values for the difference in survival time. Green ticks indicate incompatibility with null hypothesis; red indicate compatibility; gray indicate inconclusive results.*

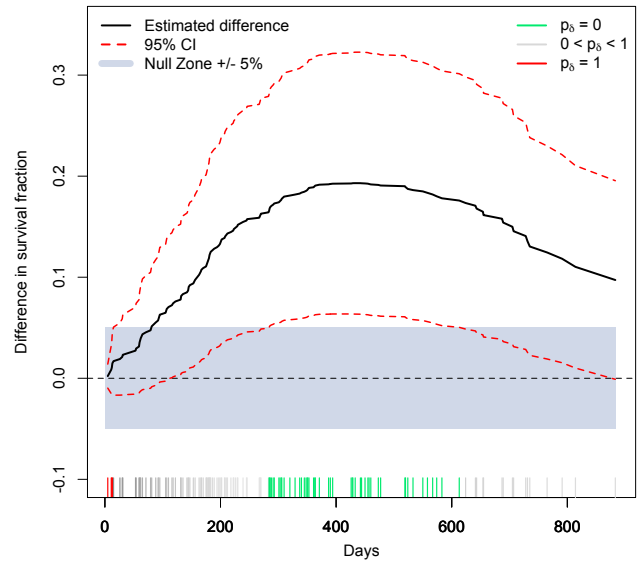

**Figure S3:** *The relationship between $P(p_\delta = 0)$ and various $\delta s$. The black line represents the traditional case when $\delta = 0$. The orange line represents $\delta = 1/30$, the green line represents $\delta = 1/2$, and the blue line represents $\delta = 1$.*

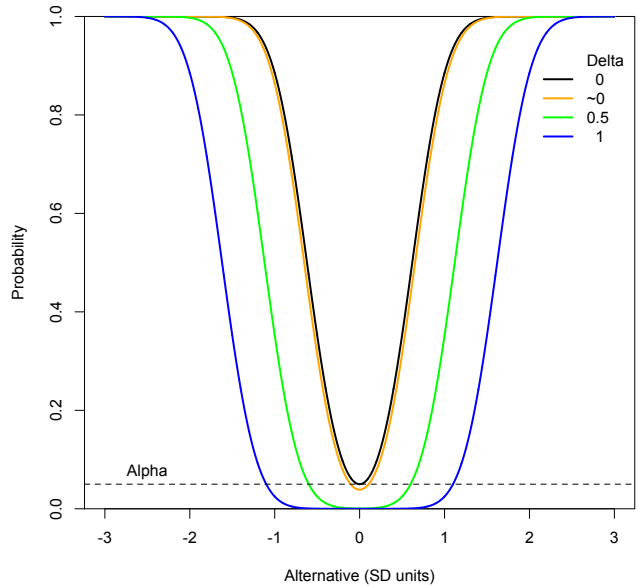

**Figure S4:** *Power curve comparisons of first-generation p-values (Black), second-generation p-values base on an indifference zone of $\delta = 0.3$ (similar to the Leukemia example), and Bonferroni adjusted p-values with k=10, 100, 7128 comparisons (red, orange, green).*

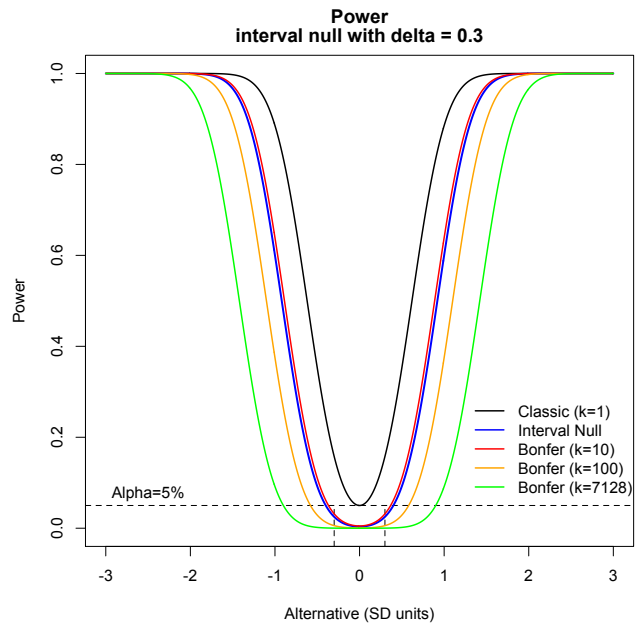

**Figure S5:** *The relationship between the probability of data supported compatibility with the null hypothesis, $P(p_\delta = 1)$, and various $\delta$s. The black line represents $\delta = 0$, the traditional point null hypothesis. The orange line represents $\delta = 1/30 \sim 0$, a very small indifference zone relative to the observed precision. The green line represents $\delta = 1/2$, and the blue line represents $\delta = 1$, which are two larger indifference zone. The graph on the left has a smaller sample size while the graph on the right has a larger sample size.*

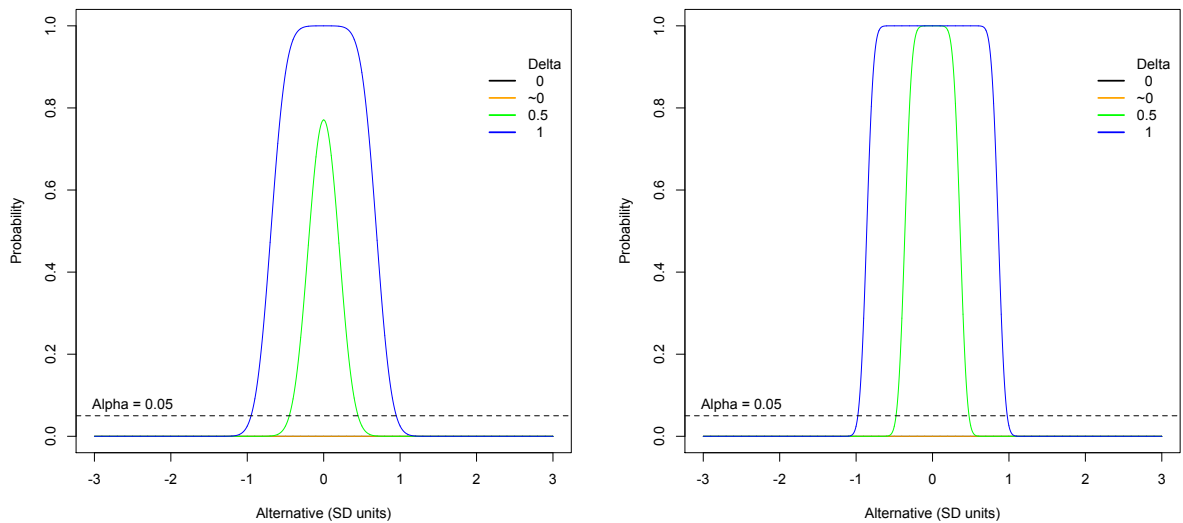

**Figure S6:** *The relationship between the probability of inconclusive results, $P(0 < p_\delta < 1)$, and various $\delta s$. The black line represents $\delta = 0.008$ which is very close to the traditional point null hypothesis. The orange line represents $\delta = 1/30 = 0.03$, a very small indifference zone relative to the observed precision. The green line represents $\delta = 1/2$, and the blue line represents $\delta = 1$, which are two larger indifference zone. The graph on the left has a smaller sample size while the graph on the right has a larger sample size.*

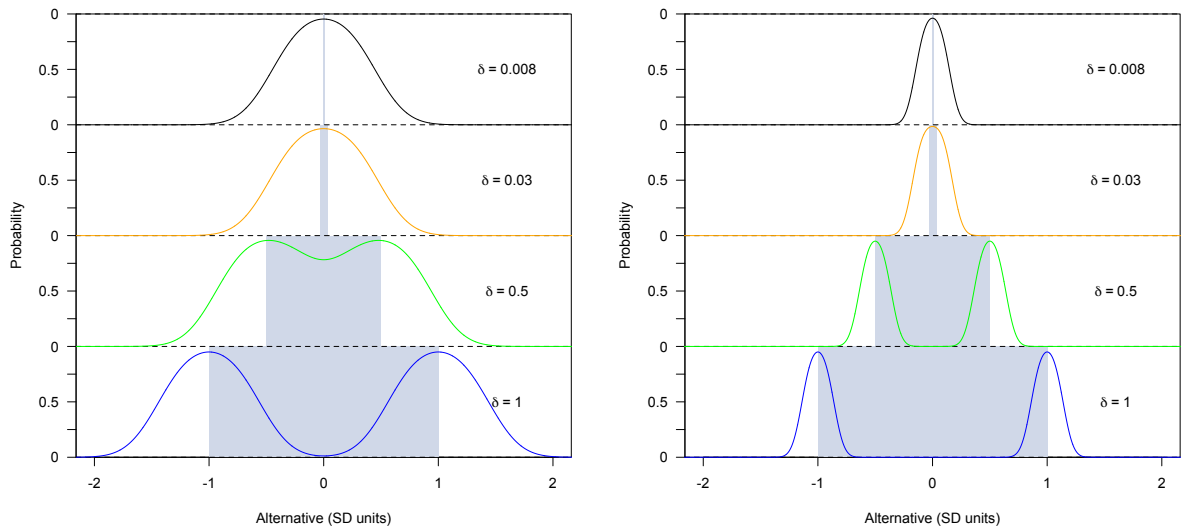

**Figure S7:** *Illustrations of the false discovery rate (red) and false confirmation rate (blue) for second-generation p-values (solid lines). The false discovery rate (red) and false non-discovery rate (blue) from a comparable hypothesis test are shown as dotted lines. This example uses* $r = 1, \alpha = 0.05, \delta = \sigma/2,$ *and* $n = 5, 20, 60, 100$.

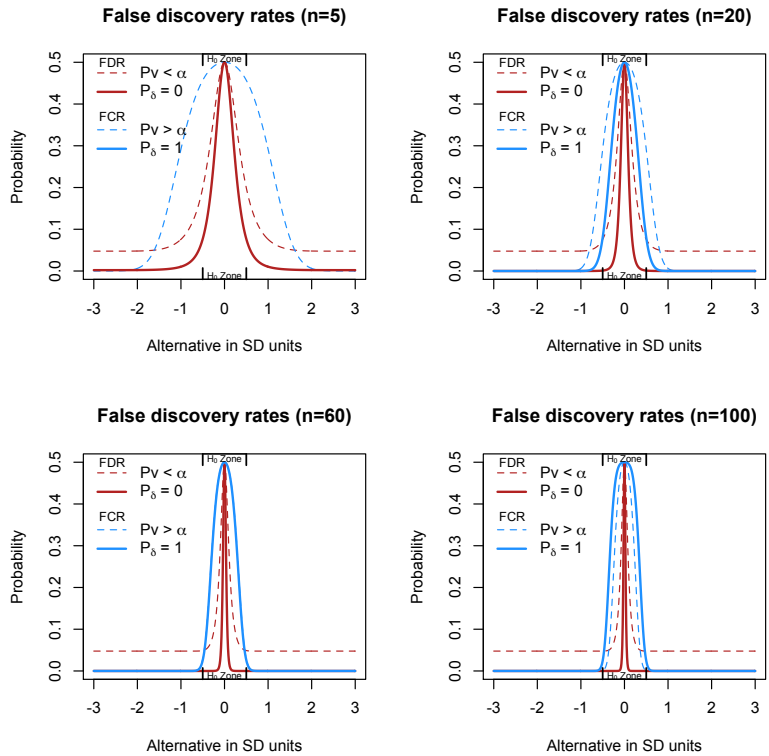